\documentstyle[12pt,aaspp4]{article}
\received{2002 October 2}
\begin{document}

\title {VISCOSITY-DRIVEN WINDS 
FROM MAGNETIZED ACCRETION DISKS}
\author {DAIZO MARUTA AND OSAMU KABURAKI}
\affil {Astronomical Institute, Graduate School of Science, 
Tohoku University, \\Aoba-ku, Sendai 980-8578, Japan;\\
maruta@astr.tohoku.ac.jp, okabu@astr.tohoku.ac.jp}

\begin{abstract}

We present an analytic model in which an inefficiently radiating 
accretion disk drives upward wind from its surfaces. 
The accretion process is controlled simultaneously by a global 
magnetic field penetrating the disk and by a viscosity of the 
accreting plasma. 
It is shown that energy is transported radially outward within 
the accreting flow, associated with the viscous angular-momentum 
transport in the same direction, and that this addition of energy 
from the inner part to the disk part makes the latter possible to 
drive an upward wind. 
The parameter that specifies the strength of a wind is determined 
uniquely in terms of a naturally introduced viscosity parameter. 

\end{abstract}


\section{INTRODUCTION}

In recent years, optically thin ADAFs (advection-dominated accretion 
flows) have been paid much attention as plausible states of 
accretion flows in the low-luminosity active galactic nuclei (LLAGNs) 
or dim galactic nuclei (DGNs), and also as states around the black
hole candidates in the Galactic binary systems (see e.g., Narayan,
Mahadevan \& Quataert 1998; Lasota 1999). 
The essence of the flows of this type is, in fact, not in the dominance 
of advection but in the inefficiency of radiation cooling. 
In this sense, they make a good contrast with the `classical' branch 
of radiatively cooled accretion flows (so called standard accretion 
disks, see Shakura \& Sunyaev 1973; Frank, King \& Raine 1992; Kato, 
Fukue \& Mineshige 1998), and hence they should be called more suitably 
as `inefficiently radiating accretion flows' (or IRAFs). 

The limit of completely advective accretion flows (i.e., ideal ADAFs) 
is realized, within the framework of IRAFs, when the locally liberated 
gravitational energy is totally advected down the stream, without 
cooled radiatively (since the flows are IRAFs) nor exchanged among 
fluid elements through, e.g., convection, conduction or viscous 
transport. 
Therefore, an ideal ADAF means an {\it adiabatic} flow within which no 
exchange of thermal energy takes place. 
Even in this case, the accretion flow is non-{\it isentropic} owing 
to the irreversible generation of entropy associated with the
dissipation of an available part of the gravitational energy. 
In general, however, there may be a non-negligible exchange of thermal 
energy among fluid elements, and this process can be a source of local 
heating or cooling that is comparable in size with the advection 
cooling. 
The importance of such possibilities have indeed been pointed out for 
the case of convection in some recent works (e.g., Narayan, 
Igumenshchev, \& Abramowicz 2000; Quataert \& Gruzinov 2000). 

It may also be interesting to see that the presence of such a 
non-adiabaticity is shown explicitly to be essential in driving 
outflows or down-flows across the surfaces of accretion disks, based 
on an analytic model of IRAFs in an ordered magnetic field (Kaburaki 
2001, hereafter referred to as K01).  
Although the explicit mechanisms of non-adiabaticity has not been 
specified there (therefore we call it the `implicit' wind model),
the model contains a parameter $n$ that specifies the deviation from 
the adiabatic flow: i.e., for $n=0$, the solution reduces to the 
adiabatic case of Kaburaki 2000 (hereafter K00) where no wind appears. 
The result is that, when $n>0$ the disk is heated as a consequence 
of some kind of non-adiabatic energy exchange and it drives an upward 
wind, and when $n<0$, the disk is cooled and subsequently it drives 
a downward wind. 
In other words, {\it a redistribution of thermal energy in the 
accreting flow is the cause of wind launching}. 

As is well known, the fluid viscosity plays an essential role in 
non-magnetic IRAF models (Ichimaru 1977; Narayan \& Yi 1994, 1995; 
Abramowicz et al. 1995; Chen et al. 1995; Blandford \& Begelman 1999). 
However, the neglection of viscosity is essential in constructing a 
completely advective (e.g., adiabatic) model of K00 because the 
viscous transport of angular momentum is inevitably accompanied by a 
certain amount of energy transport (e.g., Frank, King \& Raine 1992). 
Then, what is the effects of viscosity when it is included explicitly 
in the scheme of this otherwise-adiabatic accretion flow in a global 
magnetic field? 
Does such a flow actually drive winds, realizing an explicit example 
of the winds driven by the presence of a non-adiabaticity?
Can the value of $n$ be determined explicitly in terms of the strength 
of viscosity? 
This is what we shall investigate in the present paper. 

In the above statements (and also hereafter), we have used the words 
`winds' clearly distinguished from `jets'. 
The outflows from (or down-flows toward) the accretion disk surfaces 
not very close to the inner edge are called winds, while well 
collimated outflows from the disk inner edge are called jets. 
As described in the above, we insist that the winds are generated by 
a redistribution of thermal energy within IRAFs. However, we think 
that the problem of jet launching is still an open question and some 
qualitatively different mechanisms and boundary effects, such as the 
centrifugal barrier (e.g., Chakrabarti 1997), enhanced radiation 
cooling (non-IRAF effects) near the inner edge and magnetic 
deceleration of the accretion flows due to a piled-up poloidal field, 
should be taken into account. 
The settlement of these issues is indeed a very important but  
complicated task, and therefore is beyond the scope of the present 
paper. 

Historically, the problem of wind launching has not been treated 
clearly distinguished from the jet launching. 
A vital discussion of the centrifugal launching of winds from an 
infinitely thin, cold disk was originated by Blandford \& Payne 
(1982). 
In their self-similar solution, centrifugal winds appear when the 
inclination angle, from the vertical, of the magnetic field lines are 
larger than 30$^{\circ}$, provided that the disk is rotating with 
Keplerian velocity.
Later on, Wardle \& K\"onigl (1993) examined a more realistic vertical
structure of such a launching site within a local approximation,
allowing a finite thickness of the disk.
Meanwhile, Lubow, Papaloizou \& Pringle (1994) argued that centrifugal 
winds are unstable as long as they are cool (i.e., the sound velocity
is much smaller than the Alf\'en velocity).

There is another stream of investigations of the wind/jet launching
from accretion disks led continuously by Ferreira \& Pelletier (e.g.,
1995, and references cited there), also in which the vertical structure 
in the disk is regarded as of primary importance.
In their recent paper (Casse \& Ferreira 2000), however, the 
importance of additional heating in the disk has been stressed for 
the wind/jet launching. 

In the context of IRAFs (or specifically ADAFs), the possibility
of wind/jet launching has been acknowledged by the appearance of
positive Bernoulli sum (the total energy per unit mass of the
fluid element) in the self-similar solution of Narayan \& Yi (1994,
1995).
However, Nakamura (1998) pointed out that the positivity of the
Bernoulli sum is not a genuine property of IRAFs but its sign can also
be negative depending on the radial heat transport within the flow.
Explicit examples of the solution for the accretion flow that are
accompanied by upward flows were discussed by Blandford \& Begelman
(1999), in the absence of an ordered magnetic field.
The work of K01 belongs to this stream, in which major attention
is put on the structure and energy transport in the radial direction. 

The present paper is organized as follows. 
The outline of our implicit wind model obtained in K01 is described 
first in \S 2. 
Then for the present purpose, basic equations under the presence of 
fluid viscosity are introduced in \S 3 together with various 
assumptions to simplify them, and the resulting component equations
are discussed in \S 4. 
Assuming the same variable-separated forms as adopted in K01, we 
derive in \S 5 a set of ordinary differential equations for the radial 
parts of the relevant quantities. 
We further introduce in \S 6 three specific conditions to select an IRAF 
solution among other types. 
A new type of viscosity prescription naturally follows from this 
procedure. 
The resulting solution is described in \S 7, and the energy balance 
in this solution is discussed in \S 8. 
The results are summarized in the final section.

\section{OUTLINE OF IMPLICIT WIND MODEL}

We explain here the basic ideas and physical contents of the implicit 
wind model proposed in K01.  
A schematic drawing of the global configuration presumed in this 
model is given in Figure 1. 
An asymptotically uniform magnetic field is vertically penetrating 
the accretion disk and is twisted by the rotational motion of the plasma. 
Owing to the Maxwell stress of this twisted magnetic field, a certain 
fraction of the angular momentum of accreting plasma is carried away 
to infinity, and this fact ensures the plasma to gradually infall toward 
the central black hole.
This infall sweeps the poloidal field lines toward the center, thus
resulting in the amplification of the poloidal field in the disk. 
In this course, the gravitational energy of the infalling matter is 
converted into heat as a Joule dissipation of the current induced in 
the disk, because of a finite resistivity of the plasma. 

In a stationary state, the deformation of magnetic field is determined 
by a balance between motional dragging and diffusive 
slippage of the field lines. 
Since the solution has been obtained under the assumption of large 
magnetic Reynolds number (i.e., $\Re^2(r) \gg 1$ in the disk, except 
the region close to its inner edge where $\Re \sim 1$), the field 
deformations are also large. 
In this sense the disk can be said as weakly resistive. 
The deformed part of the magnetic field and their sources (i.e., the
electric currents in the disk) are obtained by solving a set of 
resistive MHD equations consistently with the fluid motion in the disk. 
Therefore, the presence of a large-scale seed field, which is assumed 
here to be uniform for simplicity, serves in our solution only to 
specify the boundary value at the disk's outer edge. 
However, the presence of such a seed field whose origin can be 
attributed to some distant sources is essential in order not to 
conflict with the anti-dynamo theorem (e.g., Cowing 1981). 
The results also show that the deformation in the toroidal direction 
is larger than that in the poloidal direction (i.e., 
$b_{\varphi}/b_{\rm p}\sim\Re$, where $b$ denotes the deformed part 
of the magnetic field). 

The vertical structure of the disk is maintained by a pressure balance 
between the magnetic pressure of the toroidal field, which is dominant 
outside the disk, and the gas pressure in the disk. 
Thus, the accreting plasma is magnetically confined in a geometrically 
thin disk. 
Reflecting this fact, the gas pressure and density in the disk become 
quantities of order $\Re^2$ (see, equations [49] and [58] in K01). 
In general, this balance is not a static balance in its strict sense, 
and there may be a vertical flow from the upper and lower surfaces of 
a disk. 
We call such outflows (or inflows depending on the case) {\it winds}, 
distinguished from {\it jets} that may be formed in the region within 
the inner edge of the accretion disk (also see below). 
A wind emanating from the accretion disk is not expected to form a 
relativistic jet after accelerated and collimated in the wind zone 
above the disk surfaces. Instead, the presence of winds may be 
important as a mechanism of supplying hot coronae around the disks. 

Since the wind velocity obtained in K01 is much smaller than the 
rotational velocity (by a factor of order $\sim\Delta\Re^{-1}$, 
where $\Delta\ll1$ is the half-opening angle of the disk) even at the
disk surfaces, its inertial force can hardly affect the vertical force
balance in the disk. 
However, the wind may be accelerated, by some mechanisms with which 
we do not concern in this paper, to a considerable speed after it has 
been injected from the accretion disk to the wind region outside 
the disk. 
Our main interest in the present paper (and in K01) is concentrated 
only on the launch of winds recognized within the disks. 
As for the final fate of the wind, we only expect that an upward wind 
proceeds to infinity because its total energy per unit mass (i.e., 
the Bernoulli sum in K01) is positive and hence the flow is unbound 
in the gravitational field. 

It should be emphasized that, in the solution obtained in K01, the
presence of such non-adiabaticities as discussed in the previous 
section exhibits itself solely in the {\it radial profiles} of the
quantities such as density, pressure and magnetic field components
(velocity components and temperature are not affected), but does not
in the vertical profiles. 
Namely, if the radial dependences are less-steeper (i.e., $n>0$) than 
those in the critical case (i.e., $n=0$ as in the adiabatic solution 
of K00), an upward wind appears, and if they are steeper (i.e., $n<0$), 
a downward wind appears. 
The presence of a wind can be recognized also in the resulting 
radius-dependent mass accretion rate. 

The rotational velocity is a certain fraction of the Kepler velocity 
(i.e., a reduced Keplerian rotation) because in this solution the 
radial pressure-gradient force, together with the centrifugal force, 
sustains the gravitational pull on the plasma. 
In contrast to the rotational velocity, the infall velocity is a small 
quantity of the order of $\Re^{-1}$ as far as the disk is weakly 
resistive (i.e., except for the region near the inner edge). 
As for the size of resistivity, we basically consider it to be 
specified by some anomalous transport processes caused by an MHD 
turbulence, a possible origin of which we have discussed in a separate 
paper (Kaburaki, Yamazaki \& Okuyama 2002). 
Actually, however, we specify it implicitly in terms of the magnetic 
Reynolds number $\Re$ that is regarded as a free parameter, reflecting 
the rather poor current status of our understanding on the transport 
phenomena associated with turbulence. 

The analytic solution obtained in K01 describes the physical quantities 
in the accretion disk (i.e., the solution is valid only in between 
the inner and outer edges of the disk) with also a decreasing accuracy 
toward the upper and lower surfaces of the disk. 
The latter part of this statement means that the solution contains 
a few inconsistencies of the order of $\tanh^2\eta$ in the vertical 
profiles of relevant quantities,
where $\eta\equiv(\theta-\pi/2)/\Delta$ and $\theta$ is the colatitude. 
However, these inconsistencies can be safely neglected as far as the 
physics taking place near the disk midplane are concerned. 
We consider that the ``main body'' of an accretion disk occupying near 
the midplane, rather than its surface layers, plays an essential 
role in determining the wind launching, because this part contains 
the majority of plasma in the disk and hence energetically dominant 
over the surfaces layers.
The vertical structure may probably be important in considering the
origin of jets near the inner edge. 

Further, the solution does not take care of the radiation losses. 
Nevertheless, since it has been confirmed retrospectively that the 
expected radiation flux from such a disk is negligibly small as far 
as the accretion rate is sufficiently smaller than the Eddington rate, 
the solution is self-consistent within this restriction. 

If we extrapolate our physical understanding obtained within the disk 
even to the surrounding space, we are naturally led to the following 
picture of a galactic central engine. 
The engine is essentially a hydroelectric power station in which 
the ultimate energy source is the potential energy of the accreting 
plasma in the gravitational field. 
The accretion disk is a DC dynamo of an MHD type and drives mainly 
a poloidally circulating current system, which is the cause of the 
toroidal magnetic component added to the originally vertical field 
(i.e., the twisting of the field lines). 
In the configuration shown in figure 1, the radial current is driven 
outward in the accretion disk, and a part of the current closes its 
circuit through the near wind region while another part may close 
after circulating remote regions (probably reaching the boundary of 
a `cocoon' enclosing hot winds). 
Anyway, these return currents concentrate within the polar regions 
in the upper and lower hemispheres, and finally return to the inner 
edge of the disk. 

A bipolar jet may be formed from the plasma in this polar current 
regions, because the Lorentz force due to the toroidal field always 
has both of the necessary components for collimation and for radial 
acceleration, as shown in the figure. 
The often suggested universality of the association of an accretion 
disk and a bipolar jet is thus understood naturally in terms of one 
physical entity, the {\it poloidally circulating current system}. 
It is very likely that only a small fraction of the infalling 
matter input in the disk can actually fall onto the central black 
hole, with the remaining part expelled as a bipolar jet and a wind 
from disk surfaces. 

In addition to the poloidal current system discussed above, there
appears a toroidal current in the disk, which is the cause of the 
squeezed poloidal field lines.
The toroidal current is maintained against the resistive dissipation 
by the electromotive force that results from an integration of the 
motional field along a toroidal ring. It may be worth noting here 
that this important possibility cannot help being dropped from the 
beginning in the ideal-MHD treatments. This is because, in that 
approximation, the motional field is exactly balanced by the electric 
field (i.e., $\mbox{\boldmath $E$} + (1/c)\mbox{\boldmath $v$} 
\times\mbox{\boldmath $B$} = 0$), 
and hence the toroidal component of the motional field, which can be 
non-irrotational in general, is forced to vanish in a stationary, 
axisymmetric configuration according to vanishing toroidal electric 
field (i.e., $E_{\varphi}=0$; a more general discussion of the 
limitations of the ideal-MHD approximation can be seen in Kaburaki 
1998).

\section{BASIC EQUATIONS AND SIMPLIFYING ASSUMPTIONS}

For the purpose of the present paper, we first modify the set of 
resistive-MHD equations to include the viscous force in the equation
of motion:  
\begin{eqnarray}
  \frac{D\mbox{\boldmath $v$}}{Dt} 
  = - \frac{1}{\rho}\ \mbox{\boldmath $\nabla$}p 
  - \frac{1}{4\pi\rho}\ [\mbox{\boldmath $B$}\times
  (\mbox{\boldmath $\nabla$}\times\mbox{\boldmath $B$})] 
  + \mbox{\boldmath $g$} + \frac{\mbox{\boldmath $f$}}{\rho}, 
  \label{eqn:eqmom}
\end{eqnarray}
\begin{eqnarray} 
  \frac{\partial \mbox{\boldmath $B$}}{\partial t} 
    = \mbox{\boldmath $\nabla$}\times({\bf v}\times \mbox{\boldmath $B$}) 
      + \frac{c^2}{4\pi\sigma}\ \triangle\mbox{\boldmath $B$}, 
            \qquad (\sigma = \mbox{const.}) 
  \label{eqn:eqind}
\end{eqnarray}
\begin{eqnarray} 
   \frac{\partial \rho}{\partial t} 
    + \mbox{\boldmath $\nabla$}\cdot(\rho\mbox{\boldmath $v$}) = 0, 
\end{eqnarray}
\begin{eqnarray} 
  \mbox{\boldmath $\nabla$}\cdot\mbox{\boldmath $B$} = 0, 
\end{eqnarray}
where 
\begin{eqnarray}
  \frac{D\mbox{\boldmath $v$}}{Dt} &\equiv& 
  \frac{\partial \mbox{\boldmath $v$}}{\partial t} 
  + (\mbox{\boldmath $v$}\cdot\mbox{\boldmath $\nabla$})
  \mbox{\boldmath $v$}  \nonumber\\ 
  &=& \frac{\partial\mbox{\boldmath $v$}}{\partial t} 
  - \mbox{\boldmath $v$}\times[\mbox{\boldmath $\nabla$}\times
   \mbox{\boldmath $v$}] + \frac{1}{2}\ \mbox{\boldmath $\nabla$}v^2,
\end{eqnarray}
\begin{eqnarray}
  \mbox{\boldmath $f$} &=& \frac{4}{3}\ \mbox{\boldmath $\nabla$}
   (\rho\nu\mbox{\boldmath $\nabla$}\cdot\mbox{\boldmath $v$}) 
  - \mbox{\boldmath $\nabla$}\times[\mbox{\boldmath $\nabla$}
   \times(\rho\nu\mbox{\boldmath $v$})]
  + \mbox{\boldmath $\nabla$}[\mbox{\boldmath $v$}\cdot
   \mbox{\boldmath $\nabla$}(\rho\nu)] \nonumber \\
  &+& \mbox{\boldmath $\nabla$}(\rho\nu)\times
   [\mbox{\boldmath $\nabla$}\times\mbox{\boldmath $v$}] 
  - (\mbox{\boldmath $\nabla$}\cdot\mbox{\boldmath $v$})
   \ \mbox{\boldmath $\nabla$}(\rho\nu) 
  - [\triangle(\rho\nu)]\mbox{\boldmath $v$}.
\end{eqnarray}
As usual, {\boldmath $v$}, $\rho$ and $p$ denote velocity, density 
and pressure, respectively, of a fluid, and {\boldmath $B$} and 
$\sigma$, magnetic field and electric conductivity. 
The viscous force {\boldmath $f$} is expressed in terms of the 
kinematic viscosity $\nu$ only, because of Stokes's relation.  

In the above set of equations, the number of unknowns is formally 8: 
i.e., $\mbox{\boldmath $v$}$, $\mbox{\boldmath $b$}$, $p$, $\rho$. 
On the other hand, the number of independent equations is 7, owing 
to a degeneracy in Maxwell's equations (see K00). 
The lacking equation is that of energy transport, which may in general 
be hopelessly complicated to be treated analytically. 
Therefore, it is often replaced by the assumption of polytrope, for 
simplicity. 
Since, however, there is no justification for the validity of the 
polytropic relation except for the special cases of adiabatic and 
isothermal processes, we rather let the set of equations open and 
discuss the energy balance after an explicit solution has been 
obtained. 

In addition to the above 8 unknowns, actually we do not know how to 
specify the realistic sizes of viscosity $\nu$ and electric 
conductivity $\sigma$ from some fundamental theories. 
Namely, substantial number of unknowns is 10 instead of 8. 
Therefore, we need three more equations or constraints in order to 
select a definite IRAF solution among other possible types of solutions. 

Once such a solution has been obtained, we can calculate other 
quantities of our interest from the following subsidiary equations. 
The electric field {\boldmath $E$}, current density {\boldmath $j$} 
and charge density $q$ are, respectively, calculated from 
\begin{eqnarray}
  \mbox{\boldmath $E$} &=& \frac{\mbox{\boldmath $j$}}{\sigma} 
  - \frac{1}{c}\ \mbox{\boldmath $v$}\ \times\mbox{\boldmath $B$},
  \nonumber \\
  \quad \mbox{\boldmath $j$} 
   &=& \frac{1}{4\pi}\ \mbox{\boldmath $\nabla$}\times
   \mbox{\boldmath $B$}, 
  \quad q = \frac{1}{4\pi}\ \mbox{\boldmath $\nabla$}\cdot
  \mbox{\boldmath $E$}.
\end{eqnarray}
Temperature $T$, which is assumed to be common to electrons 
and ions under the expectation of a fairly turbulent plasma state, 
is given by the ideal gas law, 
\begin{eqnarray}
        T = \frac{\bar{\mu}}{R}\ \frac{p}{\rho},
 \label{eqn:igl}
\end{eqnarray}
neglecting the radiation pressure ($\bar{\mu}$ is the mean molecular 
weight and $R$ is the gas constant). 

According to the recipe of K00 and K01, we adopt spherical polar 
coordinates ($r$, $\theta$, $\varphi$) and simplify the above 
set of equations under the following assumptions. 
Namely, the flow is (i) stationary ($\partial/\partial t = 0$) 
and (ii) axisymmetric ($\partial/\partial{\varphi} = 0$), and the disk 
is (iii) geometrically thin, (iv) weakly viscous and weakly resistive, 
and we further admit (v) the dominance of the disk midplane. 

The third assumption means that $\Delta\ll 1$, where $\Delta$ is 
the half-opening angle of the disk and is assumed to be constant. 
As stated in \S 2, the presence of the magnetic force guarantees the 
realization of such a thin disk, even in a hot IRAF situation (without 
this magnetic force the disk becomes geometrically thick; see, e.g., 
Kato, Fukue \& Mineshige 1998). 
Reflecting this localized structure, we have introduced a normalized 
angular variable $\eta = (\theta-\pi/2)/\Delta$. 
Then, it becomes clear that a differentiation with respect to $\theta$ 
gives rise to a large quantity of order $\Delta^{-1}$ 
(i.e., $\partial/\partial\theta = \Delta^{-1}\partial/\partial\eta$). 
We can also safely assume in the disk that $\sin\theta \simeq 1$ and 
$\cos \theta \simeq 0$. 

The statement (iv) implies that the ``representative'' values of 
the (viscous) Reynolds number $\Re_{\rm v}$ and the magnetic Reynolds 
number $\Re_{\rm m}$, whose precise definitions will be introduced 
later, are both large. 
We further assume that they are of the same order of magnitude so 
that $\Re^2_{\rm v} \sim \Re^2_{\rm m} \equiv \Re^2 \gg 1$ 
in the disk  (near the inner edge $r_{\rm in}$, however, 
$\Re \sim \Re_{\rm m} \sim 1$). 
This is because we are mainly interested in such a situation in the 
present paper, though in general there may be other possibilities. 
Namely, our present purpose is to examine whether the inclusion of a 
viscosity into otherwise-adiabatic accretion flow realized under the 
dominance of an ordered magnetic field can actually be altered to a 
non-adiabatic flow that can drive upward winds, or not. 

The assumption (v) states our policy in considering the wind launching. 
Since the majority of matter is concentrated around the midplane of 
an accretion disk, it is natural to expect that the physical properties 
of the flow should be controlled primarily by this main body of the disk. 
Therefore, we may seek an approximate solution that is accurate near 
the midplane even if making a sacrifice of the accuracy near the upper 
and lower surfaces of the disk. 
According to this spirit, we ignore the quantities that are proportional 
to $\tanh^2 \eta$ (except in the equation of vertical force balance 
where the vertical structure is essential), since $\tanh^2\eta \ll 1$, 
$\mbox{sech}^2\eta$ near the midplane. 
However, we do not intend to exclude the possibilities in which the 
forces that are important only near the disk surfaces can be essential 
in a different set up of the problems or in determining the subsequent 
acceleration of a wind.
Our interest here is concentrated only on the initial indication
of the wind launching. 
The important point is that there is indeed a meaningful solution 
by which we can discuss the physics of wind launching, within our 
restriction of the problems.

\section{APPROXIMATE COMPONENT EQUATIONS}

The method of obtaining the set of component expressions of resistive 
MHD equations simplified under the assumptions (i) through (v) is the 
same as described in K01. 
First we pick up only the leading order terms in $\Delta$ in each 
component equation, by regarding all quantities except $b_{\theta}$ and 
$v_{\theta}$ (which are of the order of $\Delta$ as specified by   
equations [\ref{eqn:mcns}] and [\ref{eqn:fcns}] below) are of order 
unity with respect to $\Delta$. 
Then further terms are dropped by the assumptions (iv) and (v). 
The only difference from the previous case may appear from the newly 
included term of viscous force in the equation of motion. 

We cite below the set of simplified equations in the present case. 
It contains all terms that appeared in K01, and in addition, all of 
the contributions from the newly included viscous term (which can 
be identified easily by the presence of $\nu$) within the approximation 
to the leading order in $\Delta$. 
\begin{enumerate} 
\item Equation of motion: \\
$r$-component,
 \begin{eqnarray}
   \frac{1}{\rho}\ \frac{\partial p}{\partial r} 
   - \frac{GM}{r^2}- \frac{v_{\varphi}^2}{r} 
   = \frac{1}{\rho r^2\Delta^2}\ \frac{\partial}{\partial\eta}
   \left(\rho\nu\frac{\partial v_r}{\partial\eta}\right)
 \label{eqn:EOMR}
 \end{eqnarray}
$\theta$-component,
 \begin{eqnarray} 
  \frac{\partial p}{\partial\eta} 
  + \frac{1}{8\pi}\ \frac{\partial b_{\varphi}^2}{\partial\eta} = 0
   \label{eqn:EOMth}
 \end{eqnarray}
$\varphi$-component,
 \begin{eqnarray}
  \lefteqn{\frac{v_r}{r}\frac{\partial}{\partial r} (rv_{\varphi}) 
   = \frac{b_{\theta}}{4\pi\rho r\Delta}
   \ \frac{\partial b_{\varphi}}{\partial\eta}} \nonumber\\
   & & + \frac{1}{\rho}\left[\frac{\partial}{\partial r}\left\{
   \rho\nu r\frac{\partial}{\partial r}\left(\frac{v_{\varphi}}{r} 
   \right)\right\} + 3\rho\nu\frac{\partial}{\partial r}
   \left(\frac{v_{\varphi}}{r}\right)\right]
  \label{eqn:EOMphi}
 \end{eqnarray}
\item Induction equation:\\
poloidal component,
 \begin{eqnarray}
   \frac{v_r b_{\theta}}{c} = -\frac{c}{4\pi\sigma\Delta}
   \ \frac{1}{r}\frac{\partial b_r}{\partial\eta}
 \end{eqnarray}
$\varphi$-component,
 \begin{eqnarray}
   r^2 b_r \frac{\partial}{\partial r} 
   \left(\frac{v_{\varphi}}{r}\right) 
   - \frac{\partial}{\partial r}(rv_r b_{\varphi}) 
   + \frac{c^2}{4\pi\sigma\Delta^2 r}
   \ \frac{\partial^2 b_{\varphi}}{\partial\eta^2} = 0
   \label{eqn:indph}
 \end{eqnarray}
\item Mass conservation, 
 \begin{eqnarray}
   \frac{1}{r^2}\ \frac{\partial}{\partial r}(r^2\rho v_r) 
   + \frac{1}{\Delta r}\ \frac{\partial}{\partial\eta}(\rho v_{\theta})
   = 0
  \label{eqn:mcns}
 \end{eqnarray}
\item Magnetic flux conservation, 
 \begin{eqnarray}
   \frac{1}{r^2}\ \frac{\partial}{\partial r}(r^2 b_r) 
   + \frac{1}{\Delta r}\ \frac{\partial b_{\theta}}{\partial\eta}
   = 0
  \label{eqn:fcns}
  \end{eqnarray}
\end{enumerate}

In the $r$-component of the equation of motion, the term on the 
right-hand side is a contribution from the viscous force. 
From its outlook, it may appear as a term of order $\Delta^{-2}$. 
However, one should be careful in evaluating the actual order of 
magnitude of this term because it contains the coefficient of 
viscosity $\nu$, the size of which is difficult to specify from a 
fundamental theory. 
At any rate, the representative size of this term can be written 
as $\sim \nu v_r/\Delta^2 L^2$ while that of the gravity is 
$\sim v_{\rm K}^2/L$, where $L$ is a representative length-scale 
and $v_{\rm K}$ is the Kepler velocity. 
Recalling that $v_r \sim v_{\rm K}/\Re_{\rm m}$ (K00, K01), we can 
evaluate their ratio as $(\Delta^2\Re_{\rm v}\Re_{\rm m})^{-1} 
\sim (\Delta\Re)^{-2}$, where we have introduced $\Re_{\rm v} 
\equiv Lv_{\rm K}/\nu$ and used the assumption $\Re_{\rm v}\sim 
\Re_{\rm m}\equiv \Re$. 

Thus, it turns out that the viscous term is much larger than 
the other terms (that are of order unity) as long as 
$(\Delta\Re)^{-2}\gg 1$, which will be confirmed later in \S8 though 
$\Re$ may depend on $\Delta$ in general. 
Then, this term should vanish by itself since there is no other 
term that can balance it. 
Therefore, we have  
  \begin{eqnarray} 
   \frac{\partial}{\partial\eta}
   \left(\rho\nu\frac{\partial v_r}{\partial\eta}\right) = 0,
  \label{eqn:fr}
  \end{eqnarray} 
supplemented by 
  \begin{eqnarray} 
         \frac{GM}{r^2}=\frac{v_{\varphi}^2}{r}
         -\frac{1}{\rho}\ \frac{\partial p}{\partial r}, 
  \label{eqn:EOMr}
  \end{eqnarray} 
from equation (\ref{eqn:EOMR}). 
We cannot drop the latter equation here as an equation of the 
next-leading order, because without it we cannot specify the rotational 
velocity. In this sense, we should regard rather that the former 
equation contains only a subsidiary information though it is a 
leading order equation (see also the discussion in the next section). 

Another contribution from the viscous force appears in the 
$\varphi$-component of equation of motion (\ref{eqn:EOMphi}). 
This equation describes the way in which the angular momentum 
is carried by the viscous stress as well as by the Maxwell stress. 
One can easily confirm that both terms on the right-hand side 
have the same order of magnitude, $v_{\rm K}^2/\Re L$ (where $\nu$ 
has been rewritten in terms of $\Re$), since $b_{\varphi}^2/(4\pi\rho)
\sim v_{\rm K}^2$ as seen in the solution of K01. 

The equation of vertical force balance (Eq. [\ref{eqn:EOMth}]) 
remains to be that of static balance in spite of the presence of 
a vertical flow, because the relevant inertial forces are negligibly 
small owing to its small speed even at the disk surfaces 
(i.e., $v_{\theta}\sim \Delta \Re^{-1}v_{\rm K}$).

\section{SEPARATION OF VARIABLES}

Here, we perform an approximate separation of variables for all the 
relevant quantities. 
The new factors we have to take care are the terms due to a 
finite viscosity: i.e., equation (\ref{eqn:fr}) and the second term 
on the right of equation (\ref{eqn:EOMphi}).
Since the former is a demand of the pure viscosity independent of 
the magnetic field, we first examine its implications. 
It is evident that equation (\ref{eqn:fr}) has a solution for $v_r$
that is independent of $\eta$, irrespective of the functional form 
of $\nu$. 
This means that a shear of $v_r$ in the vertical direction should 
vanish under the presence of even a weak viscosity. 
On the other hand, the purely hydromagnetic (i.e., inviscid) flows 
obtained in K00 and K01 have the vertical profile, $v_r \propto 
\mbox{sech}^2\eta$. 
These results are not compatible in a strict sense. 

However, we are seeking only an approximate solution that is accurate 
only in the main body of an accretion disk, as expressed in the
assumption (v). 
Within this approximation, the difference in the above two velocity 
profiles can be ignored, and we can regard that equation (\ref{eqn:fr}) 
has already been satisfied. 
Further, the new term in equation (\ref{eqn:EOMphi}) does
not cause any trouble if we assume that the viscosity has an angular
profile of the form $\nu \propto \mbox{sech}^2\eta$. 

Thus, we can follow the footsteps given in our previous papers and
write 
\begin{eqnarray}
   b_r(\xi, \eta) =\ \tilde{b}_r(\xi)\ \mbox{sech}^2\eta \tanh\eta, 
\end{eqnarray}
\begin{eqnarray}
   b_{\theta}(\xi, \eta) =\ \tilde{b}_{\theta}(\xi)\ \mbox{sech}^2\eta, 
\end{eqnarray}
\begin{eqnarray}
   b_{\varphi}(\xi, \eta) =-\tilde{b}_{\varphi}(\xi)\tanh\eta, 
\end{eqnarray}
\begin{eqnarray}
   v_r(\xi, \eta) =-\tilde{v}_r(\xi)\ \mbox{sech}^2\eta, 
\end{eqnarray}
\begin{eqnarray}
   v_{\theta}(\xi, \eta) = \tilde{v}_{\theta}(\xi)\tanh\eta, 
\label{eqn:vth}
\end{eqnarray}
\begin{eqnarray}
   v_{\varphi}(\xi, \eta) = \tilde{v}_{\varphi}(\xi), 
\end{eqnarray}
\begin{eqnarray}
   p(\xi, \eta) =\ \tilde{p}(\xi)\ \mbox{sech}^2\eta, 
\end{eqnarray}
\begin{equation}
   \rho(\xi, \eta) =\ \tilde{\rho}(\xi)\ \mbox{sech}^2\eta, 
\end{equation}
\begin{eqnarray}
   \nu(\xi, \eta) =\ \tilde{\nu}(\xi)\ \mbox{sech}^2\eta, 
\end{eqnarray}
\begin{eqnarray}
   T(\xi, \eta) = \tilde{T}(\xi), 
\end{eqnarray}
\begin{eqnarray}
   j_r(\xi, \eta) =-\tilde{j}_r(\xi)\ \mbox{sech}^2\eta, 
\end{eqnarray}
\begin{eqnarray}
   j_{\theta}(\xi, \eta) = \tilde{j}_{\theta}(\xi)\tanh\eta, 
\label{eqn:jth}
\end{eqnarray}
\begin{eqnarray}
   j_{\varphi}(\xi, \eta) =-\tilde{j}_{\varphi}(\xi)\ \mbox{sech}^4\eta, 
\end{eqnarray}
\begin{eqnarray}
   E_r(\xi, \eta) = \tilde{E}_r(\xi)\ \mbox{sech}^2\eta, 
\end{eqnarray}
\begin{eqnarray}
   E_{\theta}(\xi, \eta) = \tilde{E}_{\theta}(\xi)\ \mbox{sech}^2\eta
    \tanh\eta, 
\end{eqnarray}
\begin{eqnarray}
   E_{\varphi}(\xi, \eta) = \tilde{E}_{\varphi}(\xi)\ \mbox{sech}^4\eta, 
\end{eqnarray}
where the radial coordinate is normalized by a reference radius $r_0$ 
as $\xi=r/r_0$. 
In the problems of accretion in an ordered magnetic field, it is 
natural to choose the radius of disk's outer edge, $r_{\rm out}$, as 
$r_0$. 
The sign of $\tilde{v}_{\theta}(\xi, \eta)$ is chosen in such a way 
that a positive $\tilde{v}_{\theta}$ corresponds to an upward wind 
(i.e., an outflow) from the disk surfaces. 

Then, a set of ordinary differential equations for the radial parts 
follows from the set of simplified equations given in the previous 
section: that is 
\begin{eqnarray}
  \frac{\tilde{v}_{\varphi}^2}{r} = \frac{1}{\tilde{\rho}}\ 
    \frac{d\tilde{p}}{dr} + \frac{GM}{r^2},
  \label{eqn:Reomr}
\end{eqnarray}
\begin{eqnarray}
  \tilde{p} = \frac{\tilde{b}_{\varphi}^2}{8\pi}, 
  \quad (p+\frac{b_{\varphi}^2}{8\pi}=\tilde{p}),
  \label{eqn:Reomth}
\end{eqnarray}
\begin{eqnarray}
  \tilde{v}_r\frac{dl}{dr} &=& \frac{1}{4\pi\Delta}
  \ \frac{\tilde{b}_{\theta}\tilde{b}_{\varphi}}{\tilde{\rho}}
  \nonumber \\
  &-& \frac{r}{\tilde{\rho}}\left[\frac{d}{dr}\left(\tilde{\nu}
  \tilde{\rho}r \frac{d\Omega}{dr}\right) 
  + 3\tilde{\nu}\tilde{\rho}\frac{d\Omega}{dr}\right],
  \label{eqn:Reomph}
\end{eqnarray}
\begin{eqnarray}
  \tilde{v}_r\tilde{b}_{\theta} = \frac{c^2}{4\pi\sigma\Delta}\ 
    \frac{\tilde{b}_r}{r},
  \label{eqn:Rindp}
\end{eqnarray}
\begin{eqnarray}
  \left( r^2\frac{d\Omega}{dr} \right) \tilde{b}_r 
    - \frac{d}{dr}(r\tilde{v}_r\tilde{b}_{\varphi}) 
    + \frac{c^2}{2\pi\sigma\Delta^2}\ \frac{\tilde{b}_{\varphi}}{r}
    = 0,
  \label{eqn:Rindph}
\end{eqnarray}
\begin{eqnarray}
  \frac{\tilde{v}_{\theta}}{\tilde{v}_r} = \Delta \left[ r\frac{d}{dr}
  \ln(r^2\tilde{\rho}\tilde{v}_r) \right], 
  \label{eqn:Rmcns}
\end{eqnarray}
\begin{eqnarray}
  \frac{\tilde{b}_{\theta}}{\tilde{b}_r} = \frac{\Delta}{2} \left[ 
    r\frac{d}{dr}\ln(r^2\tilde{b}_r) \right],
  \label{eqn:Rfcns}
\end{eqnarray}
where we have defined $l\equiv r\tilde{v}_{\varphi}$ and $\Omega 
\equiv \tilde{v}_{\varphi}/r$. 

In addition to the above set, the subsidiary equations are reduced to 
\begin{eqnarray}
  \tilde{T} = \frac{\bar{\mu}}{R}\ \frac{\tilde{p}}{\tilde{\rho}},
  \label{eqn:RT}
\end{eqnarray}
\begin{eqnarray}
  \tilde{j}_r = \frac{c}{4\pi\Delta}\ \frac{\tilde{b}_{\varphi}}{r},
  \label{eqn:Rjr}
\end{eqnarray}
\begin{eqnarray}
  \tilde{j}_{\theta} = \frac{c}{4\pi}\ \frac{1}{r}\frac{d}{dr}
    (r\tilde{b}_{\varphi}),
  \label{eqn:Rjth}
\end{eqnarray}
\begin{eqnarray}
  \tilde{j}_{\varphi} = \frac{c}{4\pi\Delta}\ \frac{\tilde{b}_r}{r},
  \label{eqn:Rjph}
\end{eqnarray}
\begin{eqnarray}
  \tilde{E}_r = -\frac{\Delta}{2}\ \frac{d}{dr}(r\tilde{E}_{\theta}) 
  = \frac{1}{c}\left(\tilde{v}_{\varphi}\tilde{b}_{\theta} 
  - \frac{c^2}{4\pi\sigma\Delta}\ \frac{\tilde{b}_{\varphi}}{r}\right), 
\end{eqnarray}
\begin{eqnarray}
  \tilde{E}_{\theta} = \frac{1}{c}\ (\tilde{v}_r\tilde{b}_{\varphi} 
    - \tilde{v}_{\varphi}\tilde{b}_r),
\end{eqnarray}
\begin{eqnarray}
  \tilde{E}_{\varphi} = \frac{1}{c}\ \left( \tilde{v}_r
  \tilde{b}_{\theta} 
    - \frac{c^2}{4\pi\sigma\Delta}\ \frac{\tilde{b}_r}{r} \right) = 0, 
\end{eqnarray}
where $\tilde{E}_{\varphi}$ is forced to vanish owing to the assumption 
(ii) and this is actually confirmed from equation (\ref{eqn:Rindp}).

\section{SCALING CONDITIONS}

In order to select the IRAF solution from other types of solutions 
to the set of ordinary differential equations obtained in the 
previous section, we introduce three additional constraints here. 
These are called the scaling conditions, because they specify the 
radial scaling of some quantities in the manner that the gravity 
requires. 
These constraints result in a simple power-law dependence for the 
radial part of each physical quantity, which is the characteristic 
feature of our wind solutions. 

The first of them is the major condition in selecting the IRAF 
solution, which is characterized essentially by a virial-type 
temperature (note that the numerator is a quantity of $\sim T/r$): 
  \begin{eqnarray}
  \alpha &\equiv& 
     -\frac{1}{\rho}
    \frac{\partial p}{\partial r}\biggm/\frac{GM}{r^2} \nonumber \\
  &=& -\frac{1}{\tilde{\rho}}\frac{d\tilde{p}}{dr}\biggm/\frac{GM}{r^2}
    = {\rm const.}\ (<1).
  \label{eqn:IRAF1}
  \end{eqnarray}
Then, we obtain from equation (\ref{eqn:EOMr}) or (\ref{eqn:Reomr}) 
a reduced Keplerian rotation of the form 
  \begin{eqnarray} 
    v_{\varphi}=\tilde{v}_{\varphi}=(1-\alpha)^{1/2}\ v_{\rm K}(r), 
  \label{eqn:redK}
  \end{eqnarray}
where $v_{\rm K}(r)\equiv(GM/r)^{1/2}$ is the Kepler velocity. 
It should be noted also that the assumption of constant $\Delta$ 
is consistent with the appearance of a virial-type temperature 
(see, the appendix of K00). 

In a similar manner, the second condition requires that 
\begin{eqnarray}
   \beta &\equiv& 
   \frac{1}{b_r}\ \frac{\partial}{\partial r}
    (rv_r b_{\varphi}) \biggm/ r^2\frac{\partial}{\partial r} 
    \left(\frac{v_{\varphi}}{r}\right) \nonumber \\
   &=& \frac{1}{\tilde{b}_r}\ \frac{d}{dr}(r\tilde{v}_r
   \tilde{b}_{\varphi})\biggm/ r^2\ \frac{d\Omega}{dr} 
   = \mbox{const.}\ (<1), 
  \label{eqn:IRAF2}
\end{eqnarray} 
where $\Omega\equiv\tilde{v}_{\varphi}/r$ is also specified by the 
gravity through equation (\ref{eqn:redK}). 
Then, equation (\ref{eqn:indph}), or (\ref{eqn:Rindph}) reduces to 
\begin{eqnarray} 
   \tilde{b}_r  = -\frac{c^2}{2\pi(1-\beta)\sigma\Delta^2} 
     \left( r^3 \frac{d\Omega}{dr} \right)^{-1} \tilde{b}_{\varphi}. 
\end{eqnarray}
Substituting equation (\ref{eqn:redK}) for $\Omega$, we obtain 
\begin{eqnarray}
  \frac{\tilde{b}_{\varphi}}{\tilde{b}_r} 
    = \frac{3\pi(1-\beta)\Delta^2\sigma}{c^2}\ (1-\alpha)^{1/2}l_{\rm K} 
    \equiv \Re_{\rm m}(r),
  \label{eqn:tRe}
\end{eqnarray}
where $l_{\rm K}\equiv\sqrt{GMr}$. 
Since the right-hand side has a form of magnetic Reynolds number, 
we adopt this as a representative value of the magnetic Reynolds 
number that is characteristic to the present problem. 
With the aid of this definition of $\Re_{\rm m}$ in (\ref{eqn:tRe}), 
we can always replace $\sigma$ in terms of 
$\Re_0\equiv\Re_{\rm m}(r_0)$ by regarding the latter as a free parameter 
of the present model. 

The third condition requires that the fraction of the viscous 
angular-momentum transport among the total transport is a constant: 
\begin{eqnarray}
  \gamma &\equiv& 
   \frac{1}{\rho}\left[\frac{\partial}{\partial r}\left\{
   \rho\nu r\frac{\partial}{\partial r}\left(\frac{v_{\varphi}}{r} 
   \right)\right\} + 3\rho\nu\frac{\partial}{\partial r}
   \left(\frac{v_{\varphi}}{r}\right)\right]
   \biggm/ \frac{v_r}{r}\ \frac{\partial}{\partial r}(rv_{\varphi})
   \nonumber \\
   &=& -\frac{1}{\tilde{\rho}}\left[ \frac{d}{dr}\left(
   \tilde{\nu}\tilde{\rho}r\frac{d\Omega}{dr} \right) 
   + 3\tilde{\nu}\tilde{\rho} \frac{d\Omega}{dr} \right] 
   \biggm/ \frac{\tilde{v}_r}{r}\ \frac{dl}{dr} = \mbox{const.}\ (<1).
 \label{eqn:IRAF3}
\end{eqnarray}
As we shall see explicitly in the next section, this condition reduces 
to a phenomenological prescription of the viscosity $\tilde{\nu}(\xi)$ in 
terms of a free parameter $\gamma$. 
Combining the above condition (\ref{eqn:IRAF3}) with equation 
(\ref{eqn:Reomph}), we obtain 
\begin{eqnarray}
  (1-\gamma)\ \tilde{v}_r\frac{dl}{dr} 
  = \frac{\tilde{b}_{\theta}\ \tilde{b}_{\varphi}}
  {4\pi\Delta\tilde{\rho}}.
\label{eqn:trq}
\end{eqnarray}

\section{VISCOUS-WIND SOLUTION}

The IRAF-type solution that is selected under the above three scaling 
conditions will be called the viscous-wind solution. 
The procedure to obtain it is quite analogous to the case of K01, 
and rather straightforward. 
Namely, we start from the following form for the poloidal magnetic field, 
  \begin{eqnarray}
   \tilde{b}_r(\xi) = B_0\ \xi^{-(3/2-n)},
  \end{eqnarray}
where the parameter $n$ describes the deviation from the adiabatic 
case ($n=0$) in which no wind appears (K00). 

Substituting the above expression into equations (\ref{eqn:tRe}) and 
(\ref{eqn:Rfcns}), we obtain 
\begin{eqnarray}
   \tilde{b}_{\varphi}(\xi) &=& \Re_0 B_0\ \xi^{-(1-n)}
\end{eqnarray}
where 
\begin{eqnarray}
   \Re_0 &=& \frac{3\pi(1-\alpha)^{1/2}(1-\beta)\sigma\Delta^2}{c^2} 
           \ l_{{\rm K}0}, 
\end{eqnarray}
and 
  \begin{eqnarray}
    \tilde{b}_{\theta}(\xi) = \frac{2n+1}{4}\ \Delta B_0\ \xi^{-(3/2-n)},
  \end{eqnarray}
respectively. 
Then, it follows from equation (\ref{eqn:Reomth}) that 
  \begin{eqnarray}
   \tilde{p}(\xi) = \frac{\Re^2_0 B_0^2}{8\pi}\ \xi^{-2(1-n)}.
   \label{eqn:p}
  \end{eqnarray}
The subscript 0 is referred to each quantity at $r_0$. 

Using the expressions for $\tilde{b}_r$ and $\tilde{b}_{\theta}$, 
we obtain from equation (\ref{eqn:Rindp}) 
  \begin{eqnarray}
    \tilde{v}_r(\xi) = \frac{3(1-\alpha)^{1/2}(1-\beta)}{2n+1} 
    \ \frac{v_{{\rm K}0}}{\Re_0}\ \xi^{-1},
  \end{eqnarray}
and further from this result and equation (\ref{eqn:Rmcns}), 
  \begin{eqnarray}
   \tilde{v}_{\theta}(\xi) = \frac{6(1-\alpha)^{1/2}(1-\beta)n}{2n+1} 
    \ \frac{\Delta v_{{\rm K}0}}{\Re_0} \ \xi^{-1}. 
  \end{eqnarray}
The density is calculated from equation (\ref{eqn:trq}) as 
  \begin{eqnarray}
   \tilde{\rho}(\xi) 
    = \frac{(2n+1)^2}{24\pi(1-\alpha)(1-\beta)(1-\gamma)}
    \ \frac{\Re^2_0 B_0^2}{v_{{\rm K}0}^2}\ \xi^{-(1-2n)}
  \end{eqnarray}

With the above temporal expressions for various quantities, we 
can determine first $\beta$ from the second scaling condition 
(\ref{eqn:IRAF2}) and then $\alpha$ from the first scaling condition 
(\ref{eqn:IRAF1}) as 
\begin{eqnarray}
  \alpha &=& \frac{2(1-n)(1-\gamma)}{3-2(1-n)\gamma}, \\
  \beta &=& \frac{2}{3}(1-n), 
\end{eqnarray}
Therefore, the above temporal expressions finally become 
\begin{eqnarray}
  \Re(\xi) = \Re_0\ \xi^{1/2}, 
   \quad \Re_0 = \sqrt{\frac{(2n+1)^3}{3-2(1-n)\gamma}} 
   \ \frac{\pi\sigma\Delta^2 l_{{\rm K}0}}{c^2}
  \label{eqn:Re}
\end{eqnarray}
  \begin{eqnarray}
   \tilde{v}_r(\xi) = \sqrt{\frac{2n+1}{3-2(1-n)\gamma}}
   \ \frac{v_{{\rm K}0}}{\Re_0}\ \xi^{-1},
  \end{eqnarray}
  \begin{eqnarray}
   \tilde{v}_{\theta}(\xi) = 2n\sqrt{\frac{2n+1}{3-2(1-n)\gamma}} 
    \ \frac{\Delta v_{{\rm K}0}}{\Re_0} \ \xi^{-1},
  \end{eqnarray}
  \begin{eqnarray} 
    \tilde{v}_{\varphi}(\xi) 
    = \sqrt{\frac{2n+1}{3-2(1-n)\gamma}}\ v_{{\rm K}0}\ \xi^{-1/2}, 
  \label{eqn:vph}
  \end{eqnarray}
  \begin{eqnarray}
   \tilde{\rho}(\xi) 
    = \frac{3-2(1-n)\gamma}{8\pi(1-\gamma)}
    \ \frac{\Re^2_0 B_0^2}{v_{{\rm K}0}^2}\ \xi^{-(1-2n)},
  \label{eqn:Rrho}
  \end{eqnarray}
and for the subsidiary equations, 
  \begin{eqnarray}
   \tilde{T}(\xi) = \frac{1-\gamma}{3-2(1-n)\gamma} 
   \ \frac{\bar{\mu}v_{{\rm K}0}^2}{R}\ \xi^{-1},
  \end{eqnarray}
  \begin{eqnarray}
   \tilde{j}_r(\xi) = \left(\frac{c}{4\pi\Delta}\right) 
        \frac{\Re_0 B_0}{r_0}\ \xi^{-(2-n)},
  \end{eqnarray}
  \begin{eqnarray}
   \tilde{j}_{\theta}(\xi) = n \left(\frac{c}{4\pi}\right)
    \ \frac{\Re_0 B_0}{r_0}\ \xi^{-(2-n)},  
  \end{eqnarray}
  \begin{eqnarray}
   \tilde{j}_{\varphi}(\xi) = \left(\frac{c}{4\pi\Delta}\right) 
    \ \frac{B_0}{r_0}\ \xi^{-(5/2-n)}, 
  \end{eqnarray}
  \begin{eqnarray}
   \tilde{E}_r(\xi)=\tilde{E}_{\theta}(\xi)=\tilde{E}_{\varphi}(\xi)=0.
  \end{eqnarray}

The $r$-dependences of all relevant quantities are the same as in 
K01, and the parameter $\gamma$ appears only in their coefficients. 
Therefore, the allowed range for $n$ remains the same as before, 
i.e., $-1/4<n<1/2$. 
It is a conspicuous character of our wind solutions (irrespective of 
implicit or explicit type) that the radial part of each quantity are 
written in a simple power-law form, as a consequence of the scaling 
conditions. 
Since, however, the quantities depend generally on the two variables, 
$\xi$ and $\eta$, the solution is not of a self-similar type in which 
every quantity depends only on a definite combination of these 
variables. 

Finally from the third scaling condition (\ref{eqn:IRAF3}), we have  
  \begin{eqnarray}
   \tilde{\nu}(\xi)=\mbox{const.}\equiv\nu_0, 
  \end{eqnarray}
and solving the resulting expression of $\gamma$ for $\nu_0$, 
  \begin{eqnarray}
   \nu_0 = \frac{2\gamma}{3(4n+1)}\ \sqrt{\frac{2n+1}{3-2(1-n)\gamma}} 
   \ v_{{\rm K}0}\ \frac{r_0}{\Re_0}. 
  \label{eqn:nu0}
  \end{eqnarray}
Since $n$ is uniquely specified by $\gamma$ (see the next section), 
the latter relation serves as a phenomenological prescription of 
the viscosity in terms of a viscous parameter $\gamma$. 
This equation shows that $\Re_{\rm v} \equiv 
l_{{\rm K}0}/\nu_0 \sim \Re_0$ as expected, since $\gamma$ and $n$ 
are the quantities of order unity. 
Here, we can see some analogy to the $\alpha$-prescription, 
$\nu=\alpha V_{\rm S}H$ (where $V_{\rm S}$ is the sound velocity 
and $H$ is the half-thickness of a disk; Shakura \& Sunyaev 1973; 
Frank, King, \& Raine 1992) since $V_{\rm S}\sim v_{\rm K}$ in our 
solution. 
It may be even closer to the $\beta$-prescription of Duschl, 
Strittmatter \& Biermann (2000), $\nu=\beta v_{\varphi}r$ where 
$\beta$ is a parameter of order $\sim\Re_{\rm v}^{-1}$.

\section{ENERGY BALANCE}

In order to discuss in detail the energy budget in an accretion disk, 
we have to derive some more relations describing global aspects of 
the present solution. 
From the definition of the disk's inner edge 
(i.e., $\Re(\xi_{\rm in}) = 1$) and equation (\ref{eqn:Re}), we obtain 
\begin{eqnarray}
  \xi_{\rm in} = \Re_0^{-2}, 
\end{eqnarray}
and the condition $r_{\rm in}\ll r_{\rm out}$ guarantees that 
$\Re_0^{-2}\ll 1$. 
The conservation of magnetic flux applied to the field penetrating 
the disk, 
\begin{eqnarray}
  \pi r_{\rm in}^2\ \tilde{b}_r(r_{\rm in}) 
  = \int_{r_{\rm in}}^{r_{\rm out}}\tilde{b}_{\theta}\ 2\pi r\ dr,
\end{eqnarray}
yields the relation 
\begin{eqnarray}
  \Delta = \frac{\xi_{\rm in}^{n+1/2}}{1-\xi_{\rm in}^{n+1/2}} 
  \simeq \xi_{\rm in}^{n+1/2} = \Re_0^{-(2n+1)}. 
 \label{eqn:Delta}
\end{eqnarray}
In the above equation, the approximate expression holds because
$\xi_{\rm in}^{n+1/2}\ll 1$ when $-1/4<n<1/2$. 
We can show also that $(\Delta\Re_0)^{-2}=\Re_0^{4n}\gg1$ for $n>0$, 
except for $n$ very close to 0. 
Even in this exceptional case of vanishing $n$ (then we have 
$(\Delta\Re_0)^{-2}\sim 1$), equation (\ref{eqn:EOMR}) can be still 
satisfied if we simply maintain the requirements, equations 
(\ref{eqn:fr}) and (\ref{eqn:EOMr}), obtained in the case of 
non-vanishing $n$. 

Owing to the presence of vertical flows, the mass accretion rate becomes 
a function of radius like 
\begin{eqnarray}
  \dot{M}(\xi) = -\int_{-\infty}^{\infty}2\pi\rho v_r r^2\Delta\ d\eta 
  = \dot{M}_0\ \xi^{2n},
\end{eqnarray}
where 
\begin{eqnarray}
  \dot{M}_0 = \frac{\sqrt{(2n+1)[3-2(1-n)\gamma]}}{3(1-\gamma)}
  \ \frac{B_0^{\ 2}r_0^2}{\Re_0^{2n}v_{{\rm K}0}}.
  \label{eqn:Md0}
\end{eqnarray}
The latter equation can be used to express $r_0$ in terms of $B_0$ and 
$\dot{M}_0$, or $B_0$ in terms of $\dot{M}_0$ and $r_0$. 
We can introduce also the mass-ejection rate due to a part of the wind 
within an arbitrary radius $r$, by 
\begin{eqnarray}
  \dot{M}_{\rm w}(\xi) \equiv \dot{M}(\xi) - \dot{M}_{\rm in} 
   = \dot{M}_0\ (\xi^{2n} - \xi_{\rm in}^{2n}).
\end{eqnarray}
It should be noted that this expression does not include the mass
loss associated with a possible jet expected to be within the disk's
inner edge.

Since we have not used the energy equation in obtaining the 
viscous-wind solution, it is not satisfied yet. 
Rather, the requirement of energy balance determines the relation 
between $n$ and $\gamma$ as confirmed below. 
The Joule heating rate is calculated as 
\begin{eqnarray}
  \lefteqn{q_{\rm J}^{+}(\xi, \eta) = \frac{\mbox{\boldmath $j$}^2}
  {\sigma}\sim \frac{j_r^2}{\sigma}} \nonumber \\
  & & = \frac{3(1-\gamma)(2n+1)}{16\pi[3-2(1-n)\gamma]}
    \ \Re_0^{2n+1}\ \frac{GM\dot{M}_0}{r_0^4}
    \ \xi^{-2(2-n)}\mbox{sech}^4\eta, 
\end{eqnarray}
where $c^2/\sigma\Delta^2$ and $B_0^2$ have been eliminated by the aid 
of equations (\ref{eqn:Re}) and (\ref{eqn:Md0}). 
The viscous heating rate is given by 
\begin{eqnarray}
  \lefteqn{q_{\rm vis}^{+}(\xi, \eta) = \nu\rho
  \left(\xi\frac{d\Omega}{d\xi}\right)^2} \nonumber \\
  & & = \frac{9\gamma(2n+1)}{16\pi(4n+1)[3-2(1-n)\gamma]}
    \ \Re_0^{2n+1}\ \frac{GM\dot{M}_0}{r_0^4}
    \ \xi^{-2(2-n)}\mbox{sech}^4\eta, 
\end{eqnarray}
where the expression for $\nu_0$ (eq. [\ref{eqn:nu0}]) has been used. 
The advection cooling, on the other hand, is 
\begin{eqnarray}
  \lefteqn{q_{\rm adv}^{-}(\xi, \eta) \equiv 
   \mbox{\boldmath $\nabla$}\cdot(h\rho\mbox{\boldmath $v$}) 
   - (\mbox{\boldmath $v$}\cdot\mbox{\boldmath $\nabla$})\ p} \nonumber \\
  & & = \frac{1}{r^2}\frac{\partial}{\partial r}(r^2h\rho v_r) 
   + \frac{1}{r\Delta}\frac{\partial}{\partial\eta}(h\rho v_{\theta}) 
    - v_r\frac{\partial p}{\partial r} 
    - \frac{v_{\theta}}{r\Delta} \frac{\partial p}{\partial\eta} 
  \nonumber \\
  & & = \frac{3(1-\gamma)(4n+1)}{16\pi[3-2(1-n)\gamma]}
    \ \Re_0^{2n+1}\ \frac{GM\dot{M}_0}{r_0^4} 
    \ \xi^{-2(2-n)}\ \mbox{sech}^4\eta, 
\end{eqnarray}
where the last term in the second line has been dropped on account 
of the assumption (v). 

Since the radial dependences of the viscous-wind solution are 
identical with that of the implicit-wind solution (K01), the optical 
depth, for the electron scattering, in the vertical direction of the 
accretion disk is a quantity of similar orders of magnitude to 
the value given as equation (9) in Yamazaki, Kaburaki \& Kino (2002). 
There, one can see that the radiative cooling may be negligible as 
far as the mass accretion rate is sufficiently small compared with 
the Eddington rate. 
This fact has also been confirmed by the explicit calculations 
of radiation losses based on the analytic models of K00 and K01, 
respectively, in Kino, Kaburaki \& Yamazaki (200) and in Yamazaki, 
Kaburaki \& Kino (2002). 
These works show that only the synchrotron and associated inverse 
Compton losses in the region very close to the inner edge can be 
non-negligible (compared with the advection cooling), because of a 
relativistic temperature there. 

Then, the local energy balance in a stationary state should be 
\begin{eqnarray}
  q^+_{\rm J} + q^+_{\rm vis} = q^-_{\rm adv}, 
 \label{eqn:ebal}
\end{eqnarray}
except in the region very close to the inner edge. 
Substituting the above $q$'s into equation (\ref{eqn:ebal}), we obtain 
\begin{eqnarray}
  f(n) \equiv 8(1-\gamma)n^2+2(1-4\gamma)n-3\gamma = 0. 
\label{eqn:fn}
\end{eqnarray}
The quadratic function of $n$ expresses a downwardly convex parabola 
for a given $\gamma$ in the range $0\leq\gamma<1$. 
In the inviscid limit ($\gamma=0$), it has two solutions, $n=0$ and 
$n=-1/4$. 
The former corresponds to the completely advective flow of K00, and 
the latter, to the isentropic flows where $q_{\rm adv}^-=0$. 
However, the latter case should be removed from our consideration 
because $q_{\rm viss}^+$ becomes negative then. 

For a general viscous fluid with $\gamma>0$, we have $f(0)=-3\gamma<0$ 
and this means that one root of equation (\ref{eqn:fn}) is always 
positive while the other root is always negative. 
Further, the fact that $f(-1/4)=-(3/2)\gamma<0$ means that the 
negative root is smaller than $-1/4$. 
In order for the positive root remains within the range $n<1/2$, 
one should require that $f(1/2)=3(1-3\gamma)>0$. 
To summarize, we have a root of equation (\ref{eqn:fn}) within 
the range $0\leq n<1/2$ for a given viscosity parameter in the 
range $0\leq\gamma<1/3$, and the root is 
\begin{eqnarray}
  n=\frac{4\gamma-1+\sqrt{-8\gamma^2+16\gamma+1}}{8(1-\gamma)}. 
\end{eqnarray}
Therefore, a non-zero viscosity always drive an upward wind ($n>0$;
see figure 2). 

Hitherto, we have regarded the appearance of a positive $v_{\theta}$ 
as the launching of a wind, in spite of the fact that our solution is 
valid only within the main body of an accretion disk and we do not 
discuss the subsequent acceleration or deceleration. 
This is because we can guess the final fate of the vertical flow 
from the sign of the Bernoulli sum that is defined as the total energy 
per unit mass of a fluid element: 
\begin{eqnarray}
  B_{\rm e}(\xi, \eta) \equiv \frac{1}{2}v^2 - \frac{GM}{r} + h, 
\end{eqnarray}
where the terms on the right-hand side are the kinetic, gravitational 
and thermal energies ($h$ is the specific enthalpy), respectively. 
Since we adopt the ideal gas law (\ref{eqn:igl}), we have 
$h=(5/2)p/\rho$ irrespective of the assumption of polytrope. 
If this sum is negative the fluid element is bound in the 
gravitational field, but if it is positive the fluid is unbound and 
can reach infinity. 

The energy of the magnetic field should not be included in the above 
definition, because the flow does not carry around the field lines 
(in contrast to the ideal-MHD case) in a stationary state. 
In a stationary accretion process, the field acts only as a catalyst 
to convert gravitational energy into other forms (e.g., thermal energy). 
Especially in our wind solutions, the magnetic field does not even 
transport energy in terms of the MHD Poynting flux since the electric 
field vanishes there. 
This means that energy exchange among fluid elements through the 
Poynting flux does not play any role in our wind models, and hence 
the whole energy associated with the magnetic extraction of angular 
momentum is released locally within the disk as a Joule dissipation. 
In terms of the ``angular velocity'' of the magnetic field, 
$\Omega_{\rm F}$, this corresponds to the case of $\Omega_{\rm F}=0$,
and in terms of the circuit theory, this does to the case of no 
external load. 

Now, the remaining task in the present section is to confirm that 
the wind identified by a positive $n$ in the range $0<n<1/2$ actually 
have a positive Bernoulli sum. 
Since the kinetic energy is dominated by its rotational part, the sum 
is calculated to be 
\begin{eqnarray}
  B_{\rm e}(\xi, \eta) \simeq \frac{1}{2}v_{\varphi}^{\ 2} 
  - \frac{GM}{r} + h 
  = \frac{2n-(4n+1)\gamma}{2[3-2(1-n)\gamma]}
  \ v_{{\rm K}0}^{\ 2}\ \xi^{-1},
 \label{eqn:Bern}
\end{eqnarray}
which is independent of $\eta$. 
If we rewrite the coefficient there as 
$\gamma(2n+1)(1-2n)/(4n+1)[3-2(1-n)\gamma]$ by the aid of equation 
(\ref{eqn:fn}), it becomes evident that the expression on the rightmost 
side of the above equation is always positive for $0<n<1/2$ and hence 
for $0<\gamma<1/3$ (see also figure 2). 
Although $B_{\rm e}$ becomes zero in the limit of $\gamma=1/3$, it 
corresponds to the unrealistic limit of $n=1/2$ in which the density 
becomes independent of the radius.
This limit has therefore been excluded from our consideration from
the beginning.  

In general accretion flows, the Bernoulli sum is not conserved along 
stream lines. 
When the flow is adiabatic, however, it is conserved. 
This is evident from the following consideration. 
Namely, in the course of gradual infall process, a part of the 
gravitational energy of a fluid element goes to its rotational energy 
and the rest is released as thermal energy. 
If there is no mechanism of energy exchange through the boundary of 
the fluid element (i.e., the flow is adiabatic), the liberated energy 
remains within the element. 
In this case the Bernoulli sum remains the same along each stream line, 
because the result of the infall is a decrease in the gravitational 
energy that is exactly compensated by the increases in the kinetic 
and thermal energies. 

Therefore, if the accreting matter falls from infinity with marginally 
bound condition (i.e., with $B_{\rm e}=0$), it will remain so unless
some mechanism of energy transport intervenes. 
In many cases, the radiative cooling carries away a considerable 
part of energy from the fluid and it becomes bound (i.e.,$B_{\rm e}<0$). 
In IRAFs, however, this cannot be expected and hence other types 
of energy exchange such as convection, conduction or viscous transport 
controls the fate of the accretion flow. 

It has turned out in this section that the viscosity always acts as 
a net source of heating for the disk and drives upward wind from its 
surfaces. 
If we extrapolate this result to the region within the inner edge of 
the accretion disk, we can say that there should be an infalling part 
of the flow with negative Bernoulli sum, which is a direct consequence 
of the outward transport of energy by the viscosity. 
In other words, a part of the accretion flow in an IRAF state can 
really fall onto the central object only when the flow can expel the 
other part as wind or jet (a basic idea of this point can be seen in
Blandford 1999).

\section{SUMMARY AND CONCLUSION}

In order to construct an explicit example of the disk wind model 
in which a non-adiabatic exchange of energy drives the wind, we have 
included the viscosity into the otherwise adiabatic model (K00) of 
accretion flows in an ordered magnetic field. 
Under a plausible set of simplifying assumptions, we have derived 
an analytic solution to the set of resistive MHD equations among 
which the equation of motion contains the viscous force term. 
It describes an inefficiently-radiating accretion flow (IRAF), 
in which the angular momentum of the accreting plasma is extracted 
simultaneously by the Maxwell stress of the twisted magnetic field 
and by the viscous torque. 
Associated with the latter process, energy is transported outwardly 
and the resultant heating of the disk becomes the cause of an upward
wind. 

From the condition of energy balance, the wind parameter $n$ is 
uniquely specified in terms of our viscosity parameter $\gamma$. 
It has turned out that, for the allowed range of $\gamma$, $n$ is 
always positive indicating that the viscosity always drives an
upward wind. 
Although we can only infer the final fate of the wind from the sign 
of the Bernoulli sum, it indicates that the wind with positive $n$ 
within the allowed range of the model will indeed reach the infinity. 
We can also speculate that, owing to this outward energy transport 
by the viscosity, at least, a part of the flow within the disk's
inner edge becomes able to fall onto the central object with negative
Bernoulli sum. 

It is very subtle to specify the type of wind launching as centrifugal
or thermal. 
Although the addition of thermal energy to a disk is essential as
the cause of a wind launching, the force due to vertical pressure
gradient is balanced by that of the magnetic pressure gradient, within
our approximation. 
Similarly, the horizontal component of the gravitational force is 
balanced by the centrifugal force plus the radial pressure gradient, 
within the same approximation.
However, it should be noted that the wind velocity is proportional
to $n$ and $n$ increases with decreasing $\alpha$ (see figure 2). 
Since the rotational velocity increases with decreasing $\alpha$, 
this means that the upward wind appears only when the disk rotates
faster than a critical value.
Thus, we prefer the interpretation of centrifugally driven wind. 

If we may assume that the wind region extending above and below
the surfaces of an accretion disk is well described by the
ideal-MHD, we can roughly estimate the terminal velocity of a wind.
Since, then, the Bernoulli sum is conserved along the magnetic
surface on which the flow travels, we have
\begin{eqnarray}
  v_{\infty} \sim \sqrt{\frac{2n-(4n+1)\gamma}{3-2(1-n)\gamma}}
    \ v_{\rm K}(r), 
\end{eqnarray}
where $r$ denotes the footpoint radius of the magnetic surface.
In deriving the above expression, we have assumed that both
rotational and thermal energy decrease to zero at infinity. 
The coefficient $\sqrt{2\bar{B}_{\rm e}}$ is less than about 0.25
(see figure 2).

\clearpage
 
\begin{figure}

\plotone{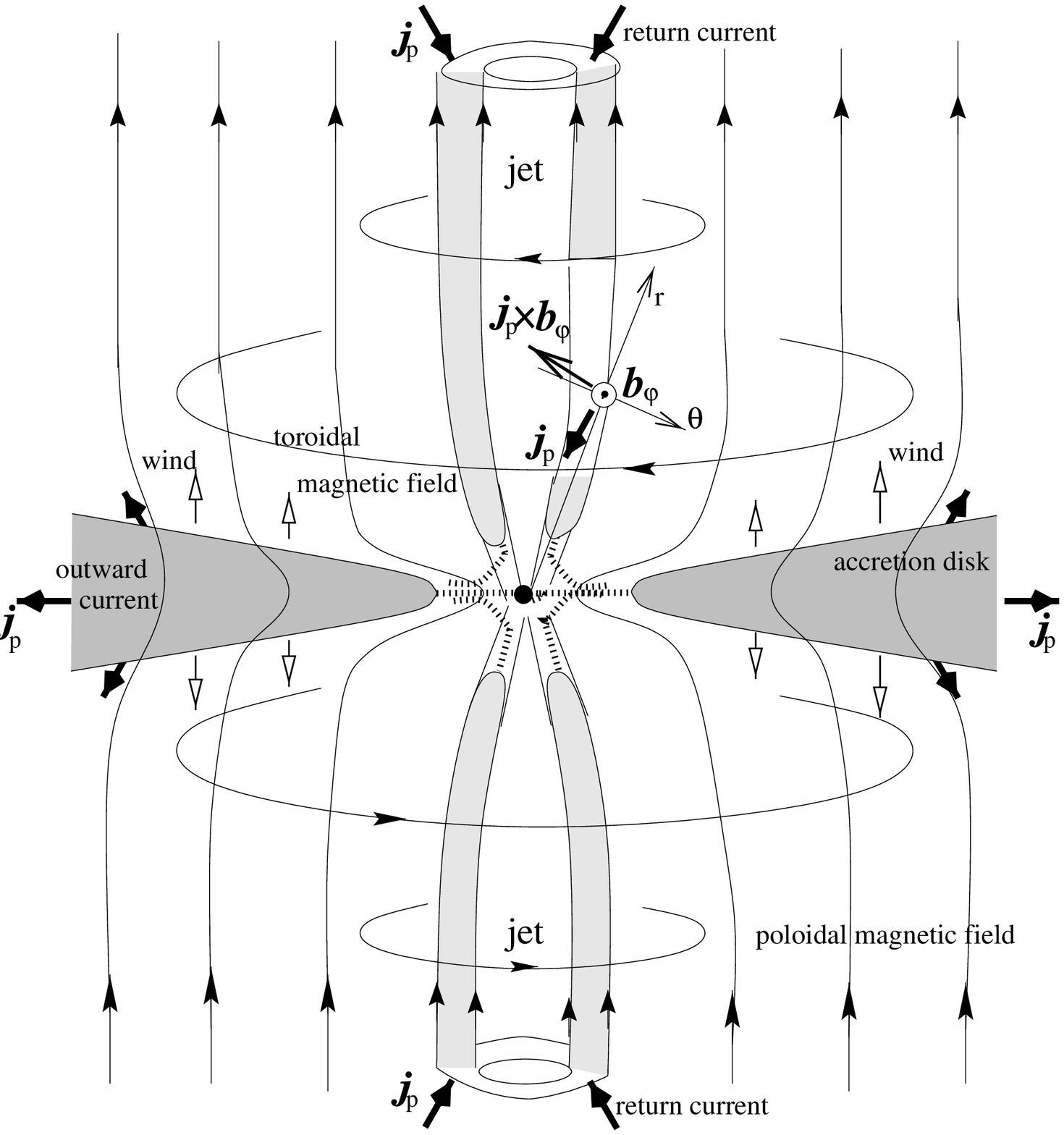}
\caption
%\figcaption  % delete this line when \begin{figure} is used 
%
{Conceptual drawing for illustration of the global configuration. 
The poloidal and toroidal components of the magnetic field lines 
are drawn with thin solid lines. 
The wind emanating from the disk surfaces and the poloidally 
circulating electric current are shown with white and black arrows, 
respectively.
Also shown is the Lorentz force acting on the return current in 
the polar region, which has the components both for collimation 
and for radial acceleration of a jet.
We distinguish jets from winds by their location and by the 
mechanisms for their generation.}
\end{figure}

\clearpage
 
\begin{figure}
\plotone{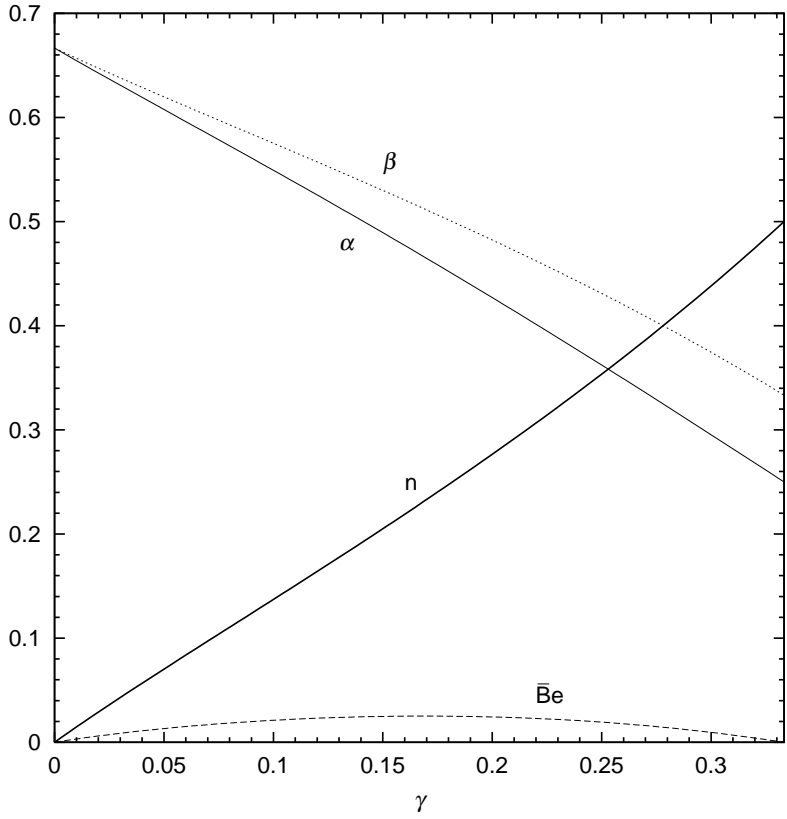}
\caption
%\figcaption  % delete this line when \begin{figure} is used 
%
{Behavior of the parameters, $\alpha$, $\beta$ and $n$, and of the
normalized Bernoulli sum, $\bar{B}_{\rm e}$, all as functions of
the viscous parameter $\gamma$.
The definition of the latter is $\bar{B}_{\rm e} \equiv
B_{\rm e}/v_{{\rm K}0}\xi^{-1}$, and it represents the coefficient
in equation (\ref{eqn:Bern}). 
The upper most value of $\gamma$ is 1/3.} 
\end{figure}

\end{document}